\documentstyle[12pt]{article}
\advance \topmargin by -\headheight
\advance \topmargin by -\headsep
     
\evensidemargin \oddsidemargin
\def\rf#1{(\ref{eq:#1})}
\def\lab#1{\label{eq:#1}}
\def\nonu{\nonumber}
\def\br{\begin{eqnarray}}
\def\er{\end{eqnarray}}
\def\be{\begin{equation}}
\def\ee{\end{equation}}
\def\eq{\!\!\!\! &=& \!\!\!\! }

\def\lb{\lbrack}
\def\rb{\rbrack}

\def\({\left(}
\def\){\right)}
\def\v{\vert}                     

\def\bc{\begin{center}}
\def\ec{\end{center}}

\newcommand\partder[2]{{{\partial {#1}}\over{\partial {#2}}}}

\newcommand\sbr[2]{\left\lbrack\,{#1}\, ,\,{#2}\,\right\rbrack} 

\def\a{\alpha}

\def\d{\delta}

\def\eps{\epsilon}

\def\l{\lambda}
\def\L{\Lambda}

\def\O{\Omega}
\def\p{\phi}
\def\P{\Phi}
\def\pa{\partial}

\def\pr{\prime}

\def\th{\theta}

\def\ti{\tilde}

\newcommand\sumi[1]{\sum_{#1}^{\infty}}   

\def\cD{{\cal D}}

\def\cKP{{\sf cKP}~}

\def\dt{{D_{\th}}}

\newcommand{\ct}[1]{\cite{#1}}
\newcommand{\bi}[1]{\bibitem{#1}}
\newcommand\NPB[3]{{\sl Nucl. Phys.} {\bf B#1} (#2) #3}

\newcommand\CMP[3]{{\sl Commun. Math. Phys.} {\bf #1} (#2) #3}

\newcommand\PLA[3]{{\sl Phys. Lett.} {\bf #1A} (#2) #3}
\newcommand\PLB[3]{{\sl Phys. Lett.} {\bf #1B} (#2) #3}
\newcommand\JMP[3]{{\sl J. Math. Phys.} {\bf #1} (#2) #3}
\newcommand\PTP[3]{{\sl Prog. Theor. Phys.} {\bf #1} (#2) #3}

\newcommand\IJMPA[3]{{\sl Int. J. Mod. Phys.} {\bf A#1} (#2) #3}

\newcommand\JPA[3]{{\sl J. Physics} {\bf A#1} (#2) #3}

\begin{document}
\begin{titlepage}
\vspace*{-1cm}
\noindent
August, 1996  \hfill{UICHEP-TH/96-15} \\
${}$ \hfill{hep-th/9608107}

\vspace {1.8cm}

\begin{center}
{\Large {\bf Manifestly Supersymmetric Lax Integrable Hierarchies}}
\end{center}
\vskip .3in
\begin{center}
{  H. Aratyn\footnotemark
\footnotetext{Work supported in part by U.S. Department of Energy,
contract DE-FG02-84ER40173} and C. Rasinariu}

\par \vskip .1in \noindent
Department of Physics \\
University of Illinois at Chicago\\
845 W. Taylor St.\\
Chicago, IL 60607-7059
\end{center}

\vspace{2cm}

\begin{abstract}

A systematic method of constructing manifestly supersymmetric
$1+1$-dimensional KP Lax hierarchies is presented.
Closed expressions for the Lax operators in terms of superfield
eigenfunctions are obtained.
All hierarchy equations being eigenfunction equations
are shown to be automatically invariant under the (extended) supersymmetry.
The supersymmetric Lax models existing in the literature are found to be
contained (up to a gauge equivalence) in our formalism.
\end{abstract}
\end{titlepage}

\underline{Introduction.}$\;\,$
In recent years, there has been rapidly growing interest in various
$1+1$-dimensional soliton systems exhibiting invariance under (extended)
supersymmetry and admitting Lax representations.
For various contributions to this program see for instance
\ct{BD,BKS,ZP,KST,KS,MP} and references therein.

Still, in our opinion there remains a need for a systematic approach
to the problem.
The purpose of this paper is to propose such a construction.
The resulting formalism is manifestly supersymmetric, encompasses known
supersymmetric models encountered in literature and provides a
straightforward guidance for the future model building.

The starting point is the SKP${}_2$ hierarchy based on the Lax operator of an
even parity and correspondingly involving only even time-flows \ct{skp2}.
The reduction scheme we propose generalizes to $N=1$ supersymmetry
the method which has previously been applied
to the standard KP Hierarchy and resulted in the constrained KP hierarchy
\ct{ckpa,ckpb}.
It is based on the symmetry constraints generated by pairs of conjugated
eigenfunctions of the original hierarchy.
It is an important feature of the formalism that the eigenfunctions
remain eigenfunctions of the reduced hierarchy \ct{ckpb}.
In the supersymmetric case the eigenfunctions are superfields and
their flows provide a complete set of hierarchy equations.
Due to the use of the manifestly covariant formalism the hierarchy equations
are all invariant under supersymmetry.
By analogy we generalize the method and obtain the $N$-extended supersymmetric
Lax hierarchy models.

We also introduce a gauge transformation to the nonstandard supersymmetric 
KP hierarchy. As in the bosonic case the gauge transformation 
induces an equivalence between standard and nonstandard Lax hierarchies.
This mapping is used to gauge some of the supersymmetric nonstandard 
Lax hierarchies present in the literature to our models.

\underline{The KP hierarchy and its reduction.}$\;\,$
The KP hierarchy consists of a set
of multi-time evolution equations in the Lax form
\be
\pa_m L \equiv \frac{\partial L}{\partial t_{m}}  = {[(L^{m})_{+} , L]},
\qquad m=1,2,3,\ldots
\lab{laxeq}
\ee
with commuting flows
\be
\frac{\partial^{2}L}{\partial t_{m}\partial t_{n}} =
\frac{\partial^{2}L}{\partial t_{n}\partial t_{m}}.
\ee
The pseudo-differential Lax operator $L$ in \rf{laxeq} is given by
\be
L = D + \sum^{\infty}_{r=-1}u_{r}D^{-r-1},~~~~~D\equiv
\frac{\partial}{\partial x}
\lab{lax1}
\ee
In eq. \rf{lax1} $(L^{m})_{+}$ denotes the purely differential part of $L^{m}$.
We adopt a notation that for any operator
$A$ and any function $f$, the symbol $A(f)$ will indicate
the action of $A$ on $f$ while the symbol $Af$ stands for the operator
product of $A$ with the zero-order (multiplication) operator $f$.
Accordingly, $D (f) = f^{\pr}$ while $D f = f^{\pr} + f D$.

Our objective is to describe the reduction of the KP hierarchy in a way which
straightforwardly generalizes to the supersymmetric case.
Our reduction procedure employs a set of eigenfunctions
and adjoint eigenfunctions which satisfy
\be
\pa_n \Phi = B_n (\Phi ) \quad; \quad
\pa_n \Psi = - B_n^{\ast} (\Psi )
\quad\;\; n=1,2, \ldots \quad; \quad B_n \equiv L^{n }_{+} \quad; \quad
B_n^{\ast} \equiv \( L^{\ast} \)^{n}_{+}
\lab{eigenlax}
\ee
for a given Lax  operator $L$ obeying Sato's flow equation
\rf{laxeq}.
In \rf{eigenlax} we have
introduced an operation of conjugation, defined by the rules
$D^{\ast} = -D$ and $(AB)^{\ast}= B^{\ast} A^{\ast}$.

The reduction procedure goes as follows.
First of all, we notice that
for a purely differential operator $A$ and arbitrary functions
$f,g$ we have the identity
\be
 \lb A\, , \, f D^{-1} g \rb_{-} =A (f) D^{-1} g-  f D^{-1} A^{\ast} (g)
\lab{tkpsi}
\ee
This together with \rf{eigenlax} gives
\be
\partder{}{t_k}\( \Phi D^{-1} \Psi\) = \lb B_k\, , \, \Phi D^{-1} \Psi\rb_{-}
\quad .
\lab{tkppsi}
\ee
We now introduce the symmetry constraints leading to the so-called
constrained KP hierarchy denoted by \cKP \ct{ckpa,ckpb}.
Let $\pa_{\a_{i}}$ be vector fields, whose action on the
standard KP Lax operator from equation \rf{lax1}
is induced by the (adjoint) eigenfunctions $\P_i, \Psi_i$ of $L$ through
\be
\pa_{\a_{i}} {L} \equiv \lb L \,  , \, \Phi_i D^{-1} \Psi_i \rb
\quad ;\quad i=1,\ldots ,m
\lab{ghostflo}
\ee
As a result of \rf{tkppsi}, the vector fields $\pa_{{\a_{i}}}$
commute with the isospectral flows of the Lax operator:
\be
\sbr{\pa_{\a_{i}}}{\pa_n} L = 0  \qquad n=1, 2, \ldots
\lab{comm}
\ee
The reduction is then performed by imposing equality between the ``ghost''
flow $\sum_{i=1}^m \pa_{{\a_{i}}}$  and the isospectral flow $\pa_r$ of the
original KP hierarchy.
Comparing \rf{ghostflo} with equation \rf{laxeq} we find that
the negative pseudo-differential part of
the Lax
operator belonging to the constrained KP hierarchy
is $L^r_{-} = \sum_{i=1}^m \Phi_i D^{-1} \Psi_i $.
The positive differential part of $L^r$ has a generic
form $L^r_{+}  = D^r+ \sum_{l=0}^{r-2} u_l D^l$.
Especially for $r=1$ we are led to the Lax operator:
\be
L = D +  \sum_{i=1}^m \Phi_i D^{-1} \Psi_i
\lab{f-5}
\ee
still subject to the Lax equation \rf{laxeq}.
For the simplest $m=1$ case the constrained system from \rf{f-5}
is equivalent to the AKNS hierarchy.
Note, that the (adjoint) eigenfunctions $\Phi_i$ and $ \Psi_i$ of the
original Lax operator $L$ \rf{lax1}
used in the above construction remain
(adjoint) eigenfunctions for $L$  \rf{f-5} \ct{ckpb}.

\underline{$N=1$ Supersymmetric KP hierarchy.}$\;\,$
Let superspace be defined in terms of the pair $(x, \th)$
with the even variable $x$ and the odd  variable $\th$.
The corresponding supercovariant derivative is
\be
\dt = \partder{}{\th} + \th D
\lab{dt}
\ee
and has the property $\dt^2 = D$.
An arbitrary pseudo-super-differential operator $\L$ has the formal expression
\begin{eqnarray}
\L = \sum_{i=-\infty}^{N}V_{i}\dt^{i} \quad.
\end{eqnarray}
We denote its splitting on positive/negative parts by
\begin{eqnarray}
\L_{+} \equiv \sum_{i=0}^{N}V_{i}\dt^{i},~~~~~\L_{-} \equiv
\sum_{i=-\infty}^{-1}V_{i}\dt^{i}.
\end{eqnarray}
The multiplication of two operators $\L$ and $\Omega$
is determined by the associativity.
The basic product rule is
$\dt UV = (\dt U)V+(-1)^{\v U\v }U\dt V$ for $\dt$ acting on
two arbitrary superfields $U$ and $V$.
The parity of a function $U$ is denoted by $\v U \v$ and is
equal to zero for $U$ being even and one for $U$ being odd.
The above identity generalizes to the supersymmetric version of the Leibniz
rule \ct{MR}:
\begin{eqnarray}
\dt^{i}U = \sum^{\infty}_{l=0}(-1)^{\v U\v (i-l)}\left[ \begin{array}{c}
i\\l
\end{array} \right] \dt^{l}(U) \dt^{i-l} \quad.
\lab{superleibniz}
\end{eqnarray}
The super-binomial coefficients
$\left[ \begin{array}{c} i\\k \end{array} \right]$ are for $i\geq 0$
\ct{MR}:
\begin{eqnarray}
\left[ \begin{array}{c}
i\\k
\end{array} \right] = \left\{ \begin{array}{ll}
0     & \mbox{for $k<0$ or $k>i$ or $(i,k)=(0,1)$ mod $2$} \\
\left( \begin{array}{c}
{[i/2]} \\ {[k/2]}
\end{array} \right)  & \mbox{for $0\leq k\leq i$ and $(i,k)\neq(0,1)$ mod $2$}
\end{array} \right. ;
\lab{superbinomial}
\end{eqnarray}
and are expressed for $i<0$ by the identity
\begin{eqnarray}
\left[ \begin{array}{c}
i\\k
\end{array} \right] = (-1)^{[k/2]}\left[ \begin{array}{c}
-i+k-1\\k
\end{array} \right].
\end{eqnarray}

There are two types of supersymmetric extensions of the KP hierarchy.
One is based on
the odd pseudo-super-differential operator \ct{MR}
\begin{eqnarray}
L_{MR} = \dt + \sum^{\infty}_{i=-1}\bar{U}_{i}
\dt^{-i-1}
\lab{mrlax}
\end{eqnarray}
and the other on the even pseudo-super-differential operator \ct{skp2}
\begin{eqnarray}
Q = \dt^{2} + \sum^{\infty}_{i=-2}U_{i}\dt^{-i-1}  ~~.
\lab{skplax}
\end{eqnarray}
The latter hierarchy is called SKP${}_2$.
It has been shown to be bi-Hamiltonian \ct{skp2}.
The coefficients $U_{i}$ in \rf{skplax} are
functions of $x,\th$ and various (even) time variables
$t_{m} (m=1,2,3, \ldots)$.
The pseudo-super-differential operator $Q$ as given by \rf{skplax}
has an even parity and accordingly $\v U_{i}\v=i+1\; {\rm mod}\, 2$.

The supersymmetric KP hierarchy involving the Lax operator $Q$ is a system of
infinitely many evolution equations for the functions $U_{i}$ following
from the Lax equations
\be
\pa_m Q \equiv \frac{\partial Q}{\partial t_{m}}  = {[(Q^{m})_{+} , Q]},
\qquad m=1,2,3,\ldots
\lab{star}
\ee
{}From now on we will only consider the supersymmetric KP hierarchy
SKP${}_2$ with the Lax operator $Q$ from \rf{skplax} and
its evolution equations \rf{star}.

We need to extend to this setting the concept of conjugation
denoted above by ``${}^{\ast}$''. The proposed rules are as follows.
On the product of two elements (including operators)
$\L$ and $\Omega$ the conjugation reverses the order and adds a phase
depending on the grade of the elements:
\be
\( \L\Omega \)^{\ast} = (-1)^{\v \L\v \v \O \v} \L^{\ast} \O^{\ast}  \quad.
\lab{conj}
\ee
Both derivatives $D$ and $\dt$ change sign under conjugation and
therefore
\be
\(D^k\)^{\ast}= (-1)^k D^k \quad;\quad
\( \dt^k\)^{\ast}= (-1)^{k(k+1)/2} \dt^k
\lab{dast}
\ee
It follows for consistency that for the superfield $U$ having an arbitrary
gradation we have $U^{\ast} = U$.

\underline{Defining reduction of SKP${}_2$.}$\;\,$
We are looking for the reduction of the original SKP${}_2$ hierarchy with
the typical grade zero Lax operator given by \rf{skplax}
and subject to the standard Lax equation \rf{star}.
For this purpose we again introduce a pair of (conjugated) eigenfunctions
$\P (x, \th),\Psi (x, \th)$ of $Q$ satisfying eqs. \rf{eigenlax} but this
time with $ B_n = \(Q^n\)_{+}$.
In complete analogy with the pure bosonic case (see \rf{tkppsi})
we have the following result:
\be
\pa_n \( \P \dt^{-1} \Psi \) = {\sbr{B_n}{\P \dt^{-1} \Psi}}_{-}
\lab{panp}
\ee
Note, that as a consequence of eqs. \rf{eigenlax} we also have
\be
\(B_n \P \dt^{-1} \Psi \)_{-} = \(B_n \P \) \dt^{-1} \Psi
   = \(\pa_n \P \) \dt^{-1} \Psi~~.
\lab{21}
\ee
On the other hand, a simple but tedious computation using
expansions for $B_n$ and $B_n^{*}$
\be
B_n=\sumi{i=0} P_i \dt^i \quad; \quad
 B_n^{*}=\sumi{k=0}(-1)^{k(k-1)/2} \dt^k P_k
\ee
yields
\be
-\( \P \dt^{-1} \Psi B_n \)_{-} =- \P \dt^{-1} B_n^{*} (\Psi)
\lab{N=1}
\ee
which along with eq. \rf{21} verifies \rf{panp}.

Let $\pa_{\a_{i}}$ be vector fields, whose action on the
standard KP Lax operator $Q$ from \rf{skplax}
is induced by
the (adjoint) eigenfunctions $\P_i, \Psi_i$ of $Q$ through 
\be
\pa_{\a_{i}} {Q} \equiv \lb Q \,  , \, \Phi_i \dt^{-1} \Psi_i \rb
\qquad i = 1 , \ldots  , m
\lab{ghostfloq}
\ee
Note, the presence of the super-covariant derivative $\dt$ in the definition
of the ``ghosts'' flows $\pa_{\a_{i}}$.

As in the bosonic case we find that the vector fields $\pa_{{\a_{i}}}$ commute
with the isospectral flows of the Lax operator $Q$:
\be
\sbr{\pa_{\a_{i}}}{\partder{}{t_n}} Q = 0  \qquad n=1, 2, \ldots
\lab{commq}
\ee
Hence following the standard procedure we define
the constrained supersymmetric KP hierarchy
via identification of the ``ghost'' flow $\sum_{i=1}^m \pa_{{\a_{i}}}$
with the isospectral flow $\pa_r$ of the
original SKP${}_2$ hierarchy.
We are in this way led to the Lax operator $Q_c = Q^r$ which for $r=1$
is given by
\be
Q_c = \dt^2  +  \sum_{i=1}^m \Phi_i \dt^{-1} \Psi_i
\lab{f-5s}
\ee

It can be shown that the
(adjoint) eigenfunctions $\Phi_i$, $ \Psi_i$ of the original Lax operator
$Q$ used in the above construction remain
(adjoint) eigenfunctions for $Q_c$  \rf{f-5s}.
Indeed the method of reference \ct{ckpb} extends to this case and one can show
that as a result of plugging $Q_c$ into eq.\rf{star} it follows that
up to a $(x,\th)$-independent phase
transformation $ \Phi\to e^{c_r}\Phi$ and $ \Psi\to e^{-c_r}\Psi$
(which leaves the Lax operator unchanged)  the superfields
$\Phi$ ($\Psi$) are (adjoint) eigenfunctions.
This observation carries over to the extended supersymmetric Lax operator
considered below.

\underline{Supersymmetric AKNS Model.}$\;\,$
As an example we now consider the constrained Lax operator constructed from 
only one pair of (adjoint) superfield eigenfunctions
\be
Q \equiv \dt^2 + \Phi \dt^{-1} \Psi
\lab{laa}
\ee
This is the supersymmetric AKNS model and we will study its corresponding
AKNS (generalized NLS) equations:
\be
\pa_t \P = (Q^2)_+ (\P) \quad ;\quad \pa_t \Psi = - (Q^2)^{\ast}_+ (\Psi)
\quad ;\quad \pa_t \equiv \pa / \pa t_2
\lab{nlsdef}
\ee
Although they follow automatically by construction they can also be
obtained by inserting $Q$ from \rf{laa} into the eq.\rf{star}.

We find:
\br
(Q^2)_+ &=& D^2 + 2 \P \( \dt \Psi \) + 2 (-1)^{\v \Psi\v} \P \Psi \dt
\lab{l2p}\\
(Q^2)^{\ast}_+ &=& D^2 + +2   (-1)^{(\v \P\v +1) (\v \Psi\v+1)}  \Psi \( \dt
\P \) + 2  (-1)^{(\v \P\v +1) (\v \Psi\v+1)} (-1)^{\v \P\v}  \Psi \P \dt
\lab{l2past}
\er
Choose now $\v \Psi \v =1$ and $\v \P \v =0$ with parametrization:
\be
\P = \p_1 + \th \psi_1 \qquad ;\qquad \Psi =  \psi_2 + \th \p_2
\lab{ppsi}
\ee
Plugging \rf{l2p}-\rf{l2past} and \rf{ppsi} into \rf{nlsdef} results in
\br
\pa_t \p_1 \eq \p_1^{\pr\pr} - 2 \p_1 \psi_2 \psi_1 + 2 \p_1^2 \p_2
\lab{nlsa}\\
\pa_t \psi_1 \eq \psi_1^{\pr\pr} + 2 \p_1 \p_1^{\pr} \psi_2 + 2 \p_1^2
\psi_2^{\pr} + 2 \p_1 \p_2 \psi_1
\lab{nlsb}\\
\pa_t \psi_2 \eq - \psi_2^{\pr\pr} -2  \psi_2 \p_1 \p_2
\lab{nlsc}\\
\pa_t \p_2 \eq - \p_2^{\pr\pr} + 2 \psi_2 \psi_1 \p_2- 2 \p_1 \p_2^2 + 2
\p_1 \psi_2 \psi_2^{\pr}
\lab{nlsd}
\er
The equations \rf{nlsa}- \rf{nlsd} contain the ordinary AKNS equations
in the limit $\psi_1, \psi_2 \to 0$.

It is straightforward to prove that the generalized AKNS equations
\rf{nlsa}-\rf{nlsd} are invariant under
the supersymmetry transformations:
\br
\d_{\eps} \p_1 &=& \eps \psi_1 \qquad ;\qquad
\d_{\eps}  \psi_1 = \eps \p_1^{\pr}
\lab{susbc}\\
\d_{\eps} \p_2 &=& \eps \psi_2^{\pr} \qquad ;\qquad
\d_{\eps}  \psi_2 = \eps \p_2
\lab{suscc}
\er
The above supersymmetry transformations can be rewritten in a following
covariant form
\be
\d_{\eps} \P = \eps \dt^{\dag} \P \quad ;\quad
\d_{\eps} \Psi = \eps \dt^{\dag} \Psi
\quad ;\quad \dt^{\dag} \equiv \partder{}{\th} - \th D
\lab{susa}
\ee
In fact, the preceding results concerning supersymmetry invariance
can be elegantly reassembled and generalized by
applying covariant notation.
As we now show the use of a covariant language will make transparent
the origin of invariance of eqs.\rf{nlsa}-\rf{nlsd} as well as
all the remaining hierarchy equations, due to their eigenfunction form
$ \pa_n \P (x, \th ) = (Q^n)_{+} \(\Phi (x, \th )\)$ and
$ \pa_n \Psi (x, \th ) = (Q^n)^{\ast}_{+} \(\Psi (x, \th )\)$.
Note, namely that in this setting applying the supersymmetry transformation
to the hierarchy
equations amounts to transforming the right hand side into
\be
\d_{\eps} (Q^n)_{+} \(\Phi\) = \eps \dt^{\dag} (Q^n)_{+} \(\Phi\)
\lab{obvious1}
\ee
since $(Q^n)_{+}$ has an expansion in terms of the superfields $\P,\Psi$.
The advantage of using covariant notation is now obvious.
We namely know that
\be
\eps \dt^{\dag} (Q^n)_{+} \(\Phi\) =  \eps \dt^{\dag} \pa_n \P
=  \pa_n  \eps \dt^{\dag} \P
\lab{obvious}
\ee
since  $\sbr{\pa_{n}}{\dt^{\dag}} =0$ for all $n$.
Hence applying $\d_{\eps} $ on both sides of
$ \pa_n \P  = (Q^n)_{+} \(\P \)$ is equivalent to acting with the derivation
$\eps \dt^{\dag}$.
Consequently, all higher flows of the constrained
supersymmetric KP hierarchy remain covariant.

\underline{Gauge Transformations for the Supersymmetric KP.}$\;\,$
In general the Lax operator can have ``constant'' terms i.e. superfields.
For the non-supersymmetric case it has been shown \ct{Kiso,rkp,swieca} that the
constant terms can be gauged away via a suitable transformation
and that the resulting Lax operator satisfies the standard Lax eq.\rf{laxeq}.
We extend this procedure to the supersymmetric case.

Let $\p$ be an arbitrary superfield.
Define
\be
G = \exp \( -\int^x \dt^{\v \p \v }\(\p \) dx\) ~~~.
\lab{g}
\ee
Then the differential operator
\be
Z = \dt^{2- \v \p \v } + \p
\lab{zdef}
\ee
can be gauge-transformed into a new operator
\be
\ti Z \equiv G^{-1} Z G
\lab{g-oper}
\ee
whose constant term is zero. We will verify this for the two possible values
of $\v \p \v $.

Let first $\v \p \v  = 0$. Then eq. \rf{g} becomes
$G = \exp \({-\int^x \p\, dx}\)$.
Using, that $\dt^2 = D$ and $G_2^{-1} G_2^{\pr} = - G_2^{-1} \p G_2$,
we obtain  as stated $ \ti Z   \equiv  G^{-1} \( D + \p \) G = D $.

Next, consider the remaining case $ \v \p \v = 1$.
Then eq. \rf{g} reads $G = \exp \( {-\int^x \dt (\p) dx} \)$
and it is straightforward to verify that indeed
$\ti Z  \equiv  G^{-1} \( \dt + \p \) G =  \dt $
due to $G^{-1} \dt \(G\) = - G^{-1} \p G $.

Consider the nonstandard Lax
$Q = D + U_{-1} + \sumi{i=0} U_i \dt^{-i-1}$
satisfying
\be
\pa_n Q = \sbr{{\(Q^n\)}_{\geq 1}}{Q}~.
\lab{ns-flow}
\ee
The next step is to investigate whether the gauge transformed of an
nonstandard Lax operator satisfies the standard SKP${}_2$ flow equations.

The claim is that if $Q$ satisfies eq. \rf{ns-flow} then $\ti Q=G^{-1 }Q G$
is a solution of the standard Lax equations \rf{star}.

Using eq. \rf{ns-flow} we obtain by direct computation
\be
\pa_n \ti Q = \sbr{G^{-1} {\(Q^n\)}_{\ge 1} G - G^{-1}\pa_n G}
              {\ti Q} ~~~.
\lab{g-lax}
\ee
In general for a pseudo-differential operator $M=\sum_{j \le m} a_j \dt^j $
, set ${\(M\)}_j = a_j \dt^j$. Then
${\(Q^n\)}_{\ge 1} = Q^n - \sum_{j \le 0} {\(Q^n\)}_j$
and consequently
$\pa_n Q = \sbr{{\(Q^n\)}_{\ge 1}}{Q}
= - \sbr{\sum_{j \le 0}{\(Q^n\)}_j}{Q}~ $,
which implies that
$$
\pa_n U_{-1} = {\(\pa_n Q\)}_0
= -{\(\sbr{\sum_{j \le 0}{\(Q^n\)}_j}{Q}\)}_0
={\({\sbr{Q}{{\(Q^n\)}_0}}\)}_0 = \partder{}{x} {\(Q^n\)}_0~~~.
$$

Therefore
$$
G^{-1} \pa_n G = -G^{-1}\( \int^x \pa_n U_{-1} dx\) G
 = -G^{-1}\(Q^n\)_0 G~~~,
$$
and eq.\rf{g-lax} becomes:
$$
\pa_n \ti Q
 = \sbr{G^{-1} {\(Q^n\)}_{\ge 1} G + G^{-1}\(Q^n\)_0 G}{\ti Q}
 = \sbr{G^{-1} \(Q^n\)_{\ge 0} G}{\ti Q}
 = \sbr{\(\ti Q\)_+}{\ti Q} \quad.
$$
In the last equality, the cancellation of the constant term in $\ti Q$
by the gauge transformation was taken into account.

\underline{Gauge connection to the Supersymmetric two-boson Hierarchy.}$\;\,$
As an example of gauge equivalence let us take the nonstandard Lax operator
of the supersymmetric two-boson hierarchy considered in \cite{BD}:
\be
L_{NS} = \dt^{2} - \dt \(\P_0 \) + \dt^{-1} \P_1
\lab{laxdas}
\ee
where $\P_0 = f_0 + \th J_0$ and $\P_1 = f_1 + \th J_1$
are fermionic superfields.  Observe that the above nonstandard Lax
operator can be transformed to the standard Lax operator from eq.\rf{laa}
via the gauge transformation
$Q = G^{-1} L_{NS} G $, where $G = \exp \( \int^{x} \dt \(\P_{0}\) dx \)$
and $Q$ is
\be
Q  = D + G^{-1} \dt^{-1} \P_1 G \quad .
\ee
This clearly suggests to make an identification
$\P \equiv G^{-1}\, ,\, \Psi \equiv \P_1 G$,
which in components reads:
\br
\p_1 = e^{-\int^x J_0 dx} \quad &;&
    \quad \psi_1 = e^{-\int^x J_0 dx} f_0
\lab{g-a} \\
\p_2 = e^{-\int^x J_0 dx} \( f_1 f_0 + J_1 \) \quad &;&
 \quad \psi_2 = e^{-\int^x J_0 dx} J_1
\lab{g-d}
\er
Thus the Lax operator \rf{laxdas} is brought to the standard form
$Q = D + \P \dt^{-1} \Psi $.
It is straightforward to show directly that indeed under
substitutions \rf{g-a}-\rf{g-d} the component equations of
reference \cite{BD} go into the AKNS eqs. \rf{nlsa} - \rf{nlsd} and
vice versa. This also follows from the general statement proved above.

\underline{$N$ Supersymmetric KP hierarchy.}$\;\,$
Let us consider the extended superspace described in terms of
$(x,\th_1,\th_2, \ldots, \th_N)$ where $x$ is even and $\th_i$ are odd
variables. Inspired by our discussion of $N=1$ case we
propose a particular form of the Lax operator as:
\be
Q = D + \P D^l \dt_1^{-1} \cdots \dt_N^{-1} \Psi ~ \qquad l < N
\lab{k-lax}
\ee
where $\P (x,\th_1,\th_2, \ldots, \th_N)$ and
$\Psi (x,\th_1,\th_2, \ldots, \th_N)$ are superfields on the $N$-extended
superspace with parity chosen such that $Q$ is even.
Denote $B_n = \(Q^n\)_{+}$. Then eq. \rf{panp} generalizes to:
\be
\pa_n \( \P D^l \dt_1^{-1} \cdots \dt_N^{-1} \Psi \) =
{\sbr{B_n}{\P D^l \dt_1^{-1} \cdots \dt_N^{-1} \Psi}}_{l,N}
\lab{k-panp}
\ee
where $\sbr{\cdot}{\cdot}_{l,N}$ denotes the commutator projected on
the pseudo-differential operator present on the l.h.s. space.
Indeed, assuming the relation \rf{eigenlax} for $\P$ and $\Psi$ and using
the expansions of $B_n$ and $B_n^*$
\br
B_n \eq \sumi{i_1,\ldots,i_N=0}A_{i_1,\ldots,i_N}\dt_1^{i_1} \cdots \dt_N^{i_N}
\lab{N-Bn} \\
B_n^{*} \eq \sumi{j_1,\ldots,j_N=0}\dt_1^{j_N} \cdots \dt_1^{j_N}
         A_{j_1,\ldots,j_N}
{\(-1\)}^{{3 \over 2}{\(\sum_{a=1}^N j_a\)}^2+{1 \over 2}\(\sum_{a=1}^N j_a \)}
\lab{N-Bnstar}
\er
we can verify directly eq.\rf{k-panp}. The expansion \rf{N-Bnstar}
takes into account that the parity of $A_{i_1,\ldots,i_N}$ is
$\v A_{i_1,\ldots,i_N}\v = i_1 + \cdots + i_N $ mod 2,
in order to assure the even character of $B_n$.

If, like before, the Lax operator $Q$ from \rf{k-lax} satisfies the
Sato's Lax hierarchy equations we find that $\P$
and $\Psi$ are (adjoint) eigenfunctions. Accordingly, their flow
equations will be manifestly invariant under supersymmetry transformations.
The above result can naturally be extended to the more general Lax operator:
\be
Q_{l,K,N}^{(m)} = D + \sum_{a=1}^m \sum_{i_1+ \cdots +i_N=K}^{i_j =0,1}
c^a_{i_1, \ldots ,i_N}  \P_a D^l \dt_1^{-i_1} \cdots \dt_N^{-i_N} \Psi_a ~,
\quad j= 1, \ldots ,N~~;~~ 0 \leq K \leq N
\lab{N-gen-lax}
\ee
where $c^a_{i_1, \ldots ,i_N}$ are constants and $ l < K$.
By analogy with $N=1$ case we call
the Lax operator written in the form \rf{N-gen-lax} the ``standard''
Lax operator.
The ``nonstandard" operator will contain constant terms.
The constant term denoted by $\p(x,\th_1,\ldots,\th_N)$ can
be gauged away via the gauge transformation
$G=\exp\(-\int^x \p(x,\th_1,\ldots,\th_N) dx \)$ as in the $N=1$ case.
Note, that there are two parameters $l$ and $K$ that specify a ``class"
of Lax operators \rf{N-gen-lax}. We cannot mix in one Lax operator
terms having different values of $l$ or $K$.

As an example, let us take the $N=2$ case. Then eq.\rf{N-gen-lax}
becomes:
\be
Q_{l,K,2}^{(m)} = D + \sum_{a=1}^m \sum_{i_1+i_2=K}^{i_j=0,1} c^a_{i_1,i_2}
 \P_a D^l \dt_1^{-i_1} \dt_2^{-i_2} \Psi_a~,\quad j=1,2 \quad ;\quad
 0 \leq K \leq 2~~.
\lab{2-lax}
\ee
Consider now the nonstandard Lax operator of supersymmetric
GNLS hierarchies proposed in \ct{BKS}
\be
 L_{NS} = D -{1 \over 2}\sum_{A=1}^M F_A(Z) {\bar F_A(Z)}
 - {1 \over 2}\sum_{A=1}^M F_A(Z) {\bar \cD} D^{-1} \cD\({\bar F_A(Z)}\)~~,
\lab{ns-gnls}
\ee
where $F_A(Z)$ and ${\bar F_A(Z)}$ is a pair of chiral and anti-chiral
superfields, $Z=(x,\th,{\bar \th})$ represents a coordinate point in $N=2$
superspace, and
\be
 \cD = \partder{}{\th} -{1 \over 2} {\bar \th} D \quad;\quad
{\bar \cD} = \partder{}{{\bar \th}} -{1 \over 2}\th D
\lab{ns-deriv}
\ee
are the supersymmetric covariant derivatives.
It is easy to see that under the change of coordinates in
superspace defined by
\be
\th_1 = {i \over 2} \( \th + {\bar \th}\) \quad;\quad
\th_2 = {1 \over 2} \(-\th + {\bar \th}\)
\lab{coord}
\ee
the supercovariant derivatives $\dt_i=\partder{}{\th_i}+\th_i D$ (i=1,2)
can be expressed as:
\be
\dt_1 = -i \( \cD + {\bar \cD} \) \quad;\quad
\dt_2 = -\cD + {\bar \cD}~~.
\lab{deriv}
\ee
Correspondingly the Lax \rf{ns-gnls} becomes:
\br
L_{NS} &=& D -{1 \over 2}\sum_{A=1}^M F_A {\bar F_A}
 +{1 \over 8}\sum_{A=1}^M \Biggl\lbrack F_A \dt_1^{-1} \dt_1\({\bar F_A}\)
\nonu \\
&-&  i F_A \dt_1^{-1} \dt_2 \({\bar F_A}\)
+ F_A \dt_2^{-1}\dt_2\({\bar F_A}\) + i F_A \dt_2^{-1}\dt_1\({\bar F_A}\)
\Biggr\rbrack
\lab{gnls}
\er
Applying the gauge transformation induced by
$G=\exp\({1 \over 2}\sum_{A=1}^M \int^x F_A {\bar F_A}\, dx \)$ the 
constant term in eq.\rf{gnls} is removed. Via the identifications
\be
\P_a =  \P_b =G^{-1} F_A \quad;\quad
\Psi_a = \dt_1 \({\bar F_A}\) G \quad ;\quad \Psi_b =\dt_2\({\bar F_A}\) 
\qquad a,b = 1, \ldots, M
\lab{ns-s}
\ee
the transformed Lax $Q=G^{-1} L_{NS} G$ can be cast into the
form \rf{2-lax} where $l=0,K=1$. Note that this form falls into the same
equivalence class $(l=0,K=1)$ with the Lax operator proposed in \ct{ZP}.

\underline{Comments.}$\;\,$
It is important to observe that although the Lax operator proposed
in eq.\rf{N-gen-lax} contains $K$ supercovariant derivatives
$\dt_i$ the corresponding hierarchy equations possess invariance under
the complete $N$-extended supersymmetry even for $K<N$.
For example, the Lax operator
$Q_{l=-1, K=0,N}^{(1)} = D +
\P (x,\th_1,\ldots, \th_N) D^{-1} \Psi (x,\th_1,\ldots, \th_N) $
will still be invariant under $N$ supersymmetry transformations
$\d_{\eps_i} \P = \eps_i \dt^{\dag}_i \P $ and
$\d_{\eps_i} \Psi = \eps_i \dt^{\dag}_i \Psi$ with $i= 1, \ldots ,N$
although the supercovariant derivatives are missing altogether
in its expression.

We would like to stress that the proposal of the Lax operator
of the $N$ extended supersymmetry was dictated by its symmetry structure
and analogy with the $N=1$ case.
It remains to supplement this proposal by investigation of the Hamiltonian
structure which we expect to underlie  the Lax hierarchy equations.
Also of interest is to investigate the possible additional symmetry structure
of the constrained KP hierarchies we proposed here along the lines of what
was done in the bosonic case in \ct{addsym}.
We plan to address these issues in the near future.

\end{document}